\documentstyle[preprint,eqsecnum,aps]{revtex}

\begin{document}
\draft \preprint{HEP/123-qed}


\title{\Large \bf Energy confinement for a relativistic magnetic
 flux tube in the ergosphere of a Kerr black hole}
\author{Vladimir S. Semenov$^{\mbox{1}}$, Sergei A. Dyadechkin$^{\mbox{1}}$,
Ivan B. Ivanov$^{\mbox{2}}$, and Helfried K. Biernat$^{\mbox{3}}$}

\address{$^{\mbox{1}}${Institute of Physics, State University, St. Petersburg, 198504
Russia, sem@geo.phys.spbu.ru}\\ $^{\mbox{2}}${Petersburg Nuclear Physics
Institute, Gatchina, 188300 Russia}\\$^{\mbox{3}}${Space Research Institute,
Austrian Academy of Sciences, Schmiedlstrasse 6, A--8042 Graz, Austria} }
\maketitle

\begin{abstract}                
\noindent{In the MHD description of plasma phenomena the concept of magnetic
field lines frozen into the plasma turns out to be very useful. We present here
a method of introducing Lagrangian coordinates into relativistic MHD equations
in general relativity, which enables a convenient mathematical formulation for
the behaviour of flux tubes. With the introduction of these Lagrangian,
so--called ``frozen--in'' coordinates, the relativistic MHD equations reduce to
a set of nonlinear 1D string equations, and the plasma may therefore be
regarded as a gas of nonlinear strings corresponding to flux tubes. Numerical
simulation shows that if such a tube/string falls
 into a Kerr black
hole, then the leading portion loses angular momentum and energy as the string
brakes, and to compensate for this loss, momentum and energy is radiated to
infinity to conserve energy and momentum for the tube. Inside the ergosphere
the energy of the leading part turns out to be negative after some time, and
the rest of the tube then gets energy from the hole. In our simulations most of
the compensated positive energy is also localized inside the ergosphere because
the inward speed of the plasma is approximately equal to the velocity of the
MHD wave which transports energy outside. Therefore, an additional physical
process has to be included which can remove energy from the ergophere. Magnetic
reconnection seems fills this role releasing Maxwellian stresses and producing
a relativistic jet.}
\end{abstract}


\narrowtext

\noindent{\bf 1. Introduction}\\

Black holes are almost certainly the central engines of quasars, active
galactic nuclei, and probably gamma ray bursts, and therefore the mechanism by
which energy is extracted from them is of great astrophysical interest. While
the exact form of this mechanism is not known, perhaps one of the best
candidate was proposed by Blanford and Znajek  \cite{1}, which in its simplest
form involves an axisymetric force-free electromagnetic configuration. The
present paper considers an alternative approach to the Blanford-Znajek process.
This is briefly described in the note \cite{2}, which is a variant of the
Penrose   mechanism \cite{3} for extracting energy from a rotating black hole.
Our argument relies on a very neat use of the analogy between the equations of
relativistic magnetohydrodynamics (RMHD) and of a gas of strings. It is a fact
that plasma embedded in a magnetic field can be considered as a collection of
flux tubes, and each flux tube turns out to behave as a non-linear string.
Therefore we can investigate the MHD flow as the time evolution of a test flux
tube similar to the test particle approach \cite{4}. Such an approach can
provide clear physical meaning to the otherwise difficult problem of analyzing
the dynamics of space plasmas (see, for example, \cite{4,5,6,7,8,9,10}.

Following these examples we shall investigate the motion of a relativistic flux
tube in the gravitational field of a Kerr black hole. The relativistic
equations for a flux tube have been discussed in the papers \cite{2,8}. Our
analysis will show that a single flux tube can extract energy from a rotating
black hole.

This paper is organize as follows: In Sections 2 and 3 we first introduce
Lagrangian coordinates in the non-relativistic MHD equations, and then extend
this to the case of general relativity. The conservation laws and gauge
condition are considered in Section 4. A qualitative scheme for a flux tube
interacting with a Kerr black hole is presented in Section 5. The numerical
algorithm and results of the calculations are described in Sections 6 and 7.
Test particle-string comparison is discussed in Section 8, and Section 9 is
devoted to the summary and discussion.\\

\noindent{\bf 2. Thin flux tube approximation}\\
\label{SR}

In order to study the time evolution of a flux tube, mathematically, we need
first of all the underlying equations. To this end we will introduce Lagrangian
coordinates into the MHD equations to obtain a convenient mathematical
formulation for the flux tube motion \cite{2}.

To illustrate the general idea we will first study the simpler non-relativistic
MHD case. The equation

\begin{eqnarray}
\rho\biggl(\frac{\partial{\mathbf{v}}}{\partial{t}}+
(\mathbf{v}\cdot\bigtriangledown)\mathbf{v}\biggl)=-\nabla\biggl(p+\frac{B^2}{8\pi}\biggl)
+\frac{1}{4\pi}(\mathbf{B}\cdot\nabla)\mathbf{B},  \label{motion}
\end{eqnarray}

describes motion of a plasma embedded in a magnetic field. Here $\rho, p,
\mathbf{v}, \mathbf{B}$ are plasma density, pressure, velocity and magnetic
field, respectively.  To distinguish a flux tube it is an advantage to rewrite
equation (\ref{motion}) in terms of  Lagrangian coordinates, using the
frozen-in condition \cite{11}:

\begin{eqnarray}
\frac{\partial}{\partial{t}}\frac{\mathbf{B}}{\rho}+
(\mathbf{v}\cdot\nabla)\frac{\mathbf{B}}{\rho}=
(\frac{\mathbf{B}}{\rho}\cdot\nabla)\mathbf{v}, \label{fzv}
\end{eqnarray}

Let us consider a small fragment $\delta r$ of a flux tube with cross section
$\delta S$. The mass of the plasma $\delta M$ as well as the magnetic flux
$\delta F_B$ in this fragment have to be conserved according to equation
(\ref{fzv}) \cite{11}:

\begin{equation}
\delta M=\rho\delta{r}\delta S=\mbox{const}, \hspace{12pt}\delta F_B=B\delta{S}
=\mbox{const}. \label{Fbm}
\end{equation}

Hence, the mass of plasma in the fragment of the tube with unit
flux $\delta \alpha$ is also conserved:
\begin{equation}
\delta \alpha=\frac{\rho\delta{r}}{B}=\mbox{const},\label{alpha}
\end{equation}

and, consequently, we can measure the length of the tube in units of mass
$\alpha$ instead of the usual Cartesian coordinates.

 As a second variable we can choose the Lagrangian time
$\tau$. It can be verified that the following relations are valid in the new,
so-called frozen-in coordinates $\tau, \alpha$ \cite{4}:

\begin{equation}
\frac{\partial{\mathbf{r}}}{\partial{\alpha}}=\frac{\mathbf{B}}{\rho},
\hspace{12pt}
\frac{\partial}{\partial{\alpha}}=\frac{\partial}{\partial{\mathbf{r}}}\cdot
\frac{\partial{\mathbf{r}}}{\partial{\mathbf{\alpha}}}=
\biggl(\frac{\mathbf{B}}{\rho}\cdot\nabla\biggl), \label{rel1}
\end{equation}

\begin{equation}
\frac{\partial{\mathbf{r}}}{\partial{\tau}} ={\mathbf{v}}, \hspace{12pt}
\frac{\partial}{\partial{\tau}}=\frac{\partial t }{\partial
\tau}+\frac{\partial}{\partial{\mathbf{r}}}\cdot
{\frac{\partial{\mathbf{r}}}{\partial \tau}}=
\frac{\partial}{\partial{t}}+(\mathbf{v}\cdot\nabla) , \label{rel2}
\end{equation}

\vspace{12pt} consequently,
\begin{equation}
\biggl(\frac{\mathbf{B}}{\rho}\cdot\nabla\biggl)\rightarrow
\frac{\partial}{\partial{\alpha}}, \label{rel3}
\end{equation}
\begin{equation}
\frac{\partial}{\partial{t}}+(\mathbf{v}\cdot\nabla)\rightarrow
\frac{\partial}{\partial{\tau}}. \label{rel4}
\end{equation}

Bearing in mind relations (\ref{rel3},\ref{rel4}) we can rewrite
the equation of motion (\ref{motion}) in terms of the frozen-in
coordinates $\tau, \alpha$:

\begin{equation}
\frac{\partial^2 \bf r }{\partial \tau^2}-\frac{1}{4\pi
}\frac{\partial}{\partial \alpha}\left(\frac{\partial({\rho \bf r
})}{\partial \alpha} \right)=
-\frac{1}{\rho}\nabla{P}(\mathbf{r}). \label{motiona}
\end{equation}

Generally speaking, the total pressure ${P}(\mathbf{r})$ is
unknown in advance, but fortunately there are a set of problems
when ${P}(\mathbf{r})$ can be considered as a given function.
Then, the plasma density $\rho$ is determined from the following
nonlinear equation:

\begin{equation}
P({\mathbf{r}})=p+\frac{B^2}{8 \pi}= c_e \rho^{\kappa}+\frac{1}{8
\pi}\rho^2\left(\frac{\partial{\mathbf{r}} }{\partial \alpha}\right)^2,
\label{P}
\end{equation}

which is just definition of the total (gas plus magnetic) pressure. Here
$\kappa$ is the polytropic index, and $c_e$ is the entropy of a fluid element
which is conserved during plasma motion:

\begin{equation}
\frac{\partial}{\partial{\tau}}\biggl(\frac{p}{\rho^\kappa}\biggl)=0.
\label{entrophy1}
\end{equation}

The MHD equation of motion (\ref{motion}), written in frozen-in coordinates
(\ref{motiona}) is the equation of a non-linear string, which is well known in
mathematical physics and plays an important role in the description of many
wave-like processes. Therefore we can find an analogy between the non-linear
string and the magnetic flux tube. Moreover, we can think of a plasma embedded
in a magnetic field as a collection of non-linear strings (flux tubes) rather
than a collection of fluid elements.

There is another way to introduce an analogy between the magnetic flux tube and
the non-linear string, using a variation  method. We can write the action
functional $S$  for a flux tube embedded in a pressure field $P({\bf r})$ as
the difference between kinetic and potential energy:

\begin{eqnarray}
S=\int\biggl[\frac{1}{2}\biggl(\frac{\partial\mathbf{r}}{\partial{\tau}}\biggl)^2
-\frac{1}{4 \pi}
\biggl(\rho\frac{\partial\mathbf{r}}{\partial\alpha}\biggl)^2-\varepsilon-
\frac{P(\mathbf{r})}{\rho}\biggl]d\alpha{d}\tau, \label{Snon}
\end{eqnarray}

where $\varepsilon$ is the internal energy of the plasma. The action
(\ref{Snon}) for a flux tube is similar to that for a non-linear string. It can
be used to derive conservation laws corresponding to a cyclic variable such as
energy or momentum conservation for a tube.

To establish a bridge between non-relativistic string equations and those in
general relativity it is advantageous to write equations (\ref{motiona}) in
terms of arbitrary coordinates $(x^i=x^1, x^2,x^3)$ with line element
$ds^2=g_{ik}dx^idx^k$ using the variational method:

\begin{eqnarray}
& & \frac{\partial^2 x^l}{\partial \tau^2}+ \Gamma^l_{ik} \frac{\partial
x^i}{\partial \tau}\frac{\partial x^k }{\partial
 \tau}- \frac{\partial }{\partial \alpha}
 \left(\frac{\rho}{4 \pi }\frac{\partial x^l }{\partial \alpha}\right)\\
 \nonumber
& & \hspace*{3cm} - \frac{\rho}{4\pi }\Gamma^l_{ik}\frac{\partial x^i
}{\partial \alpha}
 \frac{\partial x^k }{\partial \alpha}=
 -\frac{g^{il}}{\rho} \frac{\partial P}{\partial {x^i}}, \label{10}
\end{eqnarray}

where $\Gamma^l_{ik}$ is the Christoffel symbol.

Let us now examine under what circumstances a test flux tube may  represent MHD
flow as a whole. It is well known that characteristics of the string equations
(\ref{motiona}-\ref{entrophy1}) are Alfv\'enic and slow waves \cite{12},
respectively:

\begin{equation}
\frac{d\alpha}{d\tau}=\pm\sqrt{ \frac{\rho}{4\pi}},\,\
\frac{d\alpha}{d\tau}=\pm\sqrt{\frac{\kappa\rho{p}}{4\pi(2P+p(\kappa-2))}}.
 \label{AlSl}
\end{equation}

 Hence, the fast wave which is also characteristic of the general MHD system of
equations \cite{11}, is left out in the string approach. The physical reason
for this is clear: the fast wave is produced by gradients in the total
pressure, but $P({\bf r})$ is fixed in equations
(\ref{motiona}-\ref{entrophy1}). There is therefore no driving mechanism for
fast waves.

Therefore, MHD problems which involve significant variations of total pressure
such as gasdynamic-type explosion, are unlikely to be applicapble using the
string equations (\ref{motiona}-\ref{entrophy1}). But processes of accumulation
and relaxation of Maxwellian tensions can very often be described by the string
approach since the total pressure does not seem to vary appreciably. Hence,
$P({\bf r})$ may be considered as a given function or at least may be
determined by a perturbation method \cite{4,7,12}.

The full set of MHD equations is difficult to solve in general, especially for
3D time-dependent problems, but application of the string method may lead to a
considerable simplification in certain problems. In particular, for those
problems in which accumulation and relaxation of Maxwellian stresses play the
dominant role under nearly constant total pressure conditions, the general 4D
MHD problems can be reduced to investigation of the behaviour of a 1D test flux
tube/string. Such a method has been successfully applied to problems in
magnetospheric \cite{5,7,12}, solar \cite{6,9} and astrophysical \cite{2,10}
plasmas.

Here we apply this idea to the accretion of magnetized plasma on a rotating
black hole.\\

\noindent{\bf 3. Frozen-in coordinates in general relativity }\\

To derive the string equations appropriate to the theory of general relativity,
we have to introduce lagrangian co-ordinates into relativistic
magnetohydrodynamic (RMHD) equations which can be presented as follows
\cite{13}:

\begin{eqnarray}
\nabla_i\rho u^i & = & 0, \label{1}
\\[1mm]
\nabla_i T^{ik} & = & 0, \label{2}
 \\[1mm]
\nabla_i(h^i u^k-h^k u^i) & = & 0. \label{3}
\end{eqnarray}
Here equation (\ref{1}) is the continuity equation, equation (\ref{2}) is the
energy-momentum
 equation, and equation (\ref{3}) is  Maxwell's equation; $u^i$ is the
time-like vector of the 4-velocity, $u^i u_i=1$,

\begin{equation}
 h^i=\ast F^{ik}u_k \label{h}
\end{equation}

  is the space-like 4-vector of the magnetic field,
$h^i h_i<0$, $\ast F^{ik}$ is the dual tensor of the electromagnetic field, and
$T^{ij}$ is the stress-energy tensor:

\begin{eqnarray}
&T^{ij}=Q u^i u^j - P g^{ij}-\frac {1}{4\pi} h^i h^j, \label{4}
\end{eqnarray}

where

\begin{eqnarray}
P \equiv p-\frac {1}{8\pi}h^k h_k,\,\
Q \equiv p+\varepsilon-\frac {1}{4\pi}h^k h_k.
\label{11}
\end{eqnarray}

Here $p$ is the plasma pressure, $P$ is the total (plasma plus magnetic)
pressure, $\varepsilon $ is the internal energy including
$\rho c^2$, and $g_{ik}$ is the metric tensor with signature $(1,-1,-1,-1)$.

It is worth noticing that the 4-vector of the magnetic field is orthogonal to
the 4-velocity:
\begin{eqnarray}
u_i h^i=0, \label{uh}
\end{eqnarray}

 which is evident because $\ast F^{ik}$ is an antisymmetric tensor.

The previous analysis and discussion raises the question whether we can
introduce a coordinate system in such a way that the trajectory $u^i$ and
magnetic field lines $h^i$ become coordinate lines, like $\alpha$ and $\tau$ in
 non-relativistic MHD. The answer is provided by Lie analysis, which
states that a necessary and sufficient condition for two vectors ${a^i}$ and
${b^i}$ to act as coordinate lines is that they satisfy the following Lie
equation \cite{14,15}:

\begin{equation}
 {a^i \nabla_i b^k}={b^i \nabla_i a^k}. \label{Lie}
\end{equation}

Hence, we have to rewrite the frozen--in equation (\ref{3}) in the
form of the Lie equation (\ref{Lie}). Generally speaking,
$\nabla_i h^i \neq 0$, but we can find a function $q$ such that

\begin{equation}
\nabla_i q h^i = 0. \label{qh}
\end{equation}

 Then using (\ref{1}) the Maxwell equation
(\ref{3})  can be rewritten in the form of a Lie derivative:

\begin{equation}
\frac{h^i}{\rho}\nabla_i \frac{u^k}{q}=
\frac{u^i}{q}\nabla_i \frac{h^k}{\rho},\label{6}
\end{equation}

and we can therefore introduce coordinates $\tau, \alpha$ such
that:

\begin{eqnarray}
x^i_\tau \equiv \frac{\partial {x^i}}{\partial \tau} =
\frac{u^i}{q}, \,\ x^i_\alpha \equiv \frac{\partial
{x^i}}{\partial \alpha} = \frac{h^i}{\rho}
 \label{7}
\end{eqnarray}

with new coordinate vectors $u^i/q, h^i/\rho$ tracing the trajectory of a fluid
element and the magnetic field in a flux tube. This implies that we define a
magnetic flux tube as a flux tube of vector field $h^i$, and a bundle of
trajectories as
 a flux tube of vector field $u^i$. Bearing in mind the normalization
 $u^i u_i=1$ it is clear that the function $q$ obeys:
\begin{equation}
q^2 g_{ik}x^i_\tau x^k_\tau=1.\label{q}
\end{equation}

We can express the four-vectors of velocity and magnetic field in terms of
3-vectors ${\bf v},{\bf B}$:

\begin{eqnarray}
u^i & = & (\gamma,\gamma \frac{\bf v}{c}), \label{u3}
\\[1mm]
h^{i} & = & \left[\frac{\gamma ({\bf v}\cdot {\bf B})}{c},\frac{{\bf
B}}{\gamma}+\frac{\gamma ({\bf v}\cdot {\bf B})}{c^2}\right], \label{h33}
\end{eqnarray}

where $\gamma=1/\sqrt{1-v^2/c^2}$ is the Lorentz factor.

When $v<<c$ equations (\ref{u3},\ref{h33}) yield $u^i=(1,\frac{\bf v}{c}),
h^i=(0,{\bf B})$, and the Lie equation (\ref{6}) is reduced to the
non-relativistic frozen-in equation (\ref{fzv}). Therefore we can conclude that
formally the Lagrangian relativistic coordinates $\tau, \alpha$ (\ref{7}) are
the direct extension of non-relativistic frozen-in coordinates $\tau, \alpha$.
It can be shown \cite{8} that the mass coordinate $\alpha$ along the
relativistic flux tube has the sense of a mass of the plasma for a tube with
unit flux in the proper system of reference. The second coordinate $\tau$ is
not any more Lagrangian or proper time, but is just a time-like parameter which
traces the flux tube in the space-time of general relativity. The function $q$
shows how much the parameter $\tau$ differs from the proper time $t_p$ which is
evident from  comparison of the following equations:
\begin{eqnarray}
\frac{\partial {x^i}}{\partial \tau} = \frac{u^i}{q}, \,\
\frac{\partial {x^i}}{\partial t_p} = u^i. \label{taut}
\end{eqnarray}

The physical reason for this difference is that the proper time $t_p$ depends
on the fluid element. Hence, generally speaking, the condition
$t_p=\mbox{const}$ can not define the flux tube as a whole, which consists of
many fluid elements. The Lie equation (\ref{6}) shows that there exists such a
parameter $\tau$ which traces the flux tube in the space-time continuum of
general relativity. We will refer to parameter $\tau$ as the string time.

So, we have shown that ideal version of the induction equation (\ref{3})
together with the continuity equation (\ref{1}) allow us to introduce the
Lagrangian coordinates $\tau, \alpha$ into the RMHD equations. The functions
$x^i(\tau, \alpha)$ sweep the 2D worldsheet in the space-time continuum  which
consists of trajectories of the fluid elements for $\alpha=\mbox{const}$, and
the magnetic field lines for $\tau=\mbox{const}$.\\

\noindent{\bf 4. Variational method and string RMHD equations }\\

Using (\ref{7}) it is possible to derive the following analog of the
non-relativistic relations (\ref{rel3},\ref{rel4}):
\begin{equation}
\frac{u^i}{q}\nabla_i= \frac{\partial}{\partial{\tau}},
\label{rel3c}
\end{equation}
\begin{equation}
\frac{h^i}{\rho}\nabla_i= \frac{\partial}{\partial{\alpha}}. \label{rel4c}
\end{equation}
 Then, the energy-momentum equation
(\ref{2}) can be rearranged to form a set of string equations in
terms of the frozen-in coordinates:
\begin{eqnarray}
& &-\frac{\partial }{\partial \tau}
 \left(\frac{Q q}{\rho}x^l_\tau\right)-
 \frac{Q q}{\rho}\Gamma^l_{ik}x^i_\tau x^k_\tau  \nonumber \\
& & \hspace*{1cm} + \frac{\partial }{\partial \alpha}
 \left(\frac{\rho}{4 \pi q}x^l_\alpha\right)+
 \frac{\rho}{4\pi q}\Gamma^l_{ik}x^i_\alpha x^k_\alpha=
 -\frac{g^{il}}{\rho q} \frac{\partial P}{\partial {x^i}}, \label{10}
\end{eqnarray}

where $\Gamma^l_{ik}$ is the Christoffel symbol as before.

The string equations (\ref{10}) for a flux tube embedded in a gravitational
field $g_{ik}(x^i)$ and a pressure field $P(x^i)$ can also be derived from the
action \cite{8}:
\begin{eqnarray}
S=-\int L(x^i_\tau,x^i_\alpha,x^i) d\tau d\alpha= -\int
\frac{Q}{\rho}\sqrt{g_{ik}x^i_{\tau}x^k_{\tau}} d\tau d\alpha, \label{13}
\end{eqnarray}
where $L$ is the Lagrangian density.

 The action (\ref{13}) is invariant under
$\tau$-reparametrization $\tau \rightarrow \tau'(\tau)$. Therefore
the canonical Hamiltonian
\begin{equation}
 \frac{\partial L}{\partial{x^i_\tau}}x^i_\tau-L, \label{h1}
\end{equation}
is easily seen to vanish identically which is always the case for systems with
"time"-reparametrization invariance \cite{16}. This implies that equations
(\ref{10}) are not independent, and we need a gauge condition to fix
parametrization.

For simplicity we derive the gauge constraint for the particular case when
entropy is initially uniform along the tube. Since entropy is conserved in the
course of the motion \cite{13}, we can conclude that entropy does not depend on
$\alpha$ for all $\tau$. Hence, the Lagrangian (\ref{13}) does not depend
explicitly on $\alpha$, which leads to the following equation:

\begin{equation}
\frac{\partial}{\partial \alpha}\left( \frac{\partial
L}{\partial{x^i_\alpha}}x^i_\alpha-L \right)=
-\frac{\partial}{\partial \tau}\left( \frac{\partial
L}{\partial{x^i_\tau}}x^i_\alpha \right), \label{h2}
\end{equation}

and we find that:

\begin{equation}
\frac{\partial}{\partial \alpha}\left( w
\sqrt{g_{ik}x^i_{\tau}x^k_{\tau}} \right)=
\frac{\partial}{\partial \tau}\left(
\frac{Q}{\sqrt{g_{ik}x^i_{\tau}x^k_{\tau}}}g_{ik}x^i_{\tau}x^i_\alpha
\right). \label{h3}
\end{equation}

Here $w=\varepsilon+p/\rho$ is the enthalpy of the plasma.

Four vectors of velocity and magnetic field are orthogonal
(\ref{uh}), hence the right hand side of the equation (\ref{h3})
has to vanish which immediately yields:

\begin{equation}
 w \sqrt{g_{ik}x^i_{\tau}x^k_{\tau}}= f(\tau), \label{h4}
\end{equation}

where $f(\tau)$ is an arbitrary function. It is appropriate to choose
parametrization $\tau \rightarrow \tau'(\tau)$ so that $f(\tau)=1$, which
corresponds to the following gauge condition:

\begin{equation}
q=\sqrt{g_{ik}x^i_{\tau}x^k_{\tau}}= \frac{1}{w}. \label{h5}
\end{equation}

Somewhat surprisingly it turns out that the function $q$ which is responsible
for the string time is defined by the plasma enthalpy.

Using the gauge condition (\ref{h5}) it can be shown that equations (\ref{10})
are of hyperbolic type with relativistic alfv\'enic characteristics

\begin{equation}
\frac{d\alpha}{d\tau}=\pm\frac{\rho}{w\sqrt{4\pi Q}},
 \label{Al}
\end{equation}

and slow mode characteristics

\begin{equation}
\frac{d\alpha}{d\tau}=\pm\sqrt{\frac{\kappa\rho{p}}{4\pi{w}^3(2
P+p(\kappa-2))}}
 \label{Sl}
\end{equation}

similar to the non-relativistic case (\ref{AlSl}).

Ones again, by formulating the RMHD equations in terms of  frozen-in
coordinates, the energy-momentum equation (\ref{2}) reduces to a set of 1D wave
equations (\ref{10}), which are in fact nonlinear string equations. The
behaviour of flux tubes can therefore be studied through solving a set of
string equations, and an analogy can be drawn between the behaviour of strings
and flux tubes \cite{2,4,7,8,9,10}.

 For a cyclic variable $x^m$ there is a conservation law which
states that:
\begin{eqnarray}
\frac{\partial}{\partial \tau}\left( \frac{\partial} {\partial
x^m_\tau}(\frac{Q}{q \rho})\right)= -\frac{\partial}{\partial
\alpha}\left(\frac{\partial} {\partial x^m_\alpha}(\frac{Q}{q
\rho})\right). \label{14}
\end{eqnarray}

In the non-relativistic limit $v^2<<c^2$ the string equations (\ref{10}), the
action (\ref{13}), and the equations of characteristics (\ref{Al},\ref{Sl}) are
reduced to their analogous  equations (\ref{motiona}), (\ref{Snon}), and
(\ref{AlSl}), respectively.\\

\noindent{\bf 5. Motion of a flux tube near  a Kerr black hole}\\

 The Kerr metric in Boyer-Lindquist coordinates is given by the
following line element \cite{14}:
\begin{eqnarray}
ds^2 & =& (1-\frac{2 M r}{\Sigma})dt^2- \frac{\Sigma}{\Delta}dr^2-\Sigma
d\theta^2 \nonumber\\ & & - (r^2+a^2+\frac{2 M r
a^2}{\Sigma}\sin^2\theta)\sin^2\theta \, d\varphi^2 \nonumber\\[1mm] & & +
\frac{4 M r a}{\Sigma} \sin^2\theta \, d\varphi \, dt, \label{15}
\end{eqnarray}
where
\begin{eqnarray}
\Delta=r^2-2 M r +a^2, \,\ \Sigma=r^2+a^2\cos^2\theta. \label{16}
\end{eqnarray}
Here $M$ and $a$ are the mass and angular momentum of the hole,
respectively, and we have used a system of units in which $c=1, \
\ G=1$.

Let us now consider a test flux tube which falls from infinity into a Kerr
black hole. For cyclic variables $t$ and $\varphi$, the energy and angular
momentum conservation laws for the flux tube can be written as:
\begin{eqnarray}
\int_{\alpha_1}^{\alpha_2} \frac{Q}{w\rho}
(g_{tt}t_\tau+g_{t\varphi}\varphi_\tau) d \alpha & = & E, \label{21}
\\
\int_{\alpha_1}^{\alpha_2} \frac{Q}{w\rho} (g_{t\varphi}t_\tau+g_{\varphi
\varphi}\varphi_\tau) d \alpha & = & -L,
 \label{22}
\end{eqnarray}
if there is no flux of energy and angular momentum through the
ends ${\alpha_1},{\alpha_2}$ of the flux tube.

It is worth  comparing the tube conservation laws (\ref{21}, \ref{22}) with
those for a test particle \cite{14}:

\begin{eqnarray}
g_{tt}t_\tau+g_{t\varphi}\varphi_\tau & = & E_p, \label{21a}
\\
g_{t\varphi}t_\tau+g_{\varphi \varphi}\varphi_\tau & = & -L_p.
 \label{22a}
\end{eqnarray}

$E_p$ and $L_p$ differ from energy and angular momentum densities of the string
in equations (\ref{21}, \ref{22}) by just factor $Q/w\rho$, i.e., a fluid
element can be considered as a particle.

Let us suppose to start off with that each element of the tube has no angular
momentum at infinity. Hence, if there is no magnetic field, no tube element
would interact with any other element, and as it falls into the black hole it
would rotate with an angular velocity
$\omega_0=-g_{t\varphi}/g_{\varphi\varphi}$ of fiducial observers which follows
from equation (\ref{22a}) \cite{14}.

Because of inhomogeneous rotation of the space around the black
hole, the flux tube becomes stretched and twisted, and
gravitational energy is partially converted into magnetic energy
of the swirling tube. The strong magnetic field (string) evidently
slows down rotation of the flux tube as it falls, hence this part
of the tube with $\Omega_{tube}=\varphi_\tau/t_\tau < \omega_0$
must have negative momentum with respect to the rotation of the
hole. On the other hand, angular momentum of the tube as a whole
needs to be conserved (\ref{22}), and to compensate positive
angular momentum has to be radiated to infinity.

This is a continuous process: the deeper the tube falls in the
hole, the more the tube is stretched, the stronger the magnetic
field gets, the slower the tube rotates, the more negative
momentum is generated, and the more positive momentum escapes to
infinity.

Now we have to take into account the fact that the stretching and twisting of
the falling flux tube is most pronounced close to, and especially inside, the
ergosphere where $g_{tt}<0$ and the energy of a particle or tube element can be
negative \cite{3,14}.

As was pointed out previously, as the flux tube falls into the hole, the
leading part progressively loses more and more angular momentum and energy due
to string braking, and by the time it encounters the ergosphere boundary it
must have a substantial amount of negative momentum and, hence, this leading
part can have negative energy inside the ergosphere.

 Since the energy of the tube as a whole has to be conserved
(\ref{21}), the part of the tube with positive energy ends up with an energy
greater than the initial tube energy, $E$. To some extent this is similar to
the Penrose  process \cite{3}, but now we do not need to invoke the interaction
or decay of particles or tubes, since just a single tube can extract energy
from the hole.\\

\noindent{\bf 6. Numerical method}\\

The treatment of the previous section was based just on the energy and momentum
conservation laws which can give only a rough description of the flux tube
motion. The relativistic string is a highly nonlinear object, hence analytical
methods can hardly provide necessary details. It seems that the only way out is
the direct numerical simulation of a test flux tube motion in the vicinity of a
rotating black hole.

To implement the numerical approach some preliminary work is needed.

{\bf 6.1 Normalization}. We normalize plasma density to its initial value
$\rho_0$, length to the radius of the ergosphere in the equatorial plane $r_g$,
velocity to the light speed $c$, time scale to the ratio $r_g/c$, magnetic
field to $\sqrt{4\pi \rho_0}c$, the plasma pressure and the energy density to
$\rho_0 c^2$, $\alpha$ to  $r_g \sqrt{\rho_0/(4\pi c^2)}$. Note that radius of
the event horizon is equal to 1/2 for an extreme rotating black hole.

{\bf 6.2 Total pressure}. The distribution  of the total pressure in the
vicinity of a black hole seems to be mainly determined by the gravitational
field. Since $g_{ik}$ is axisymmetric we may expect that the function
$P(x^i)=P(r,\theta)$ is also axisymmetric. Thus, we may expect significant
increase of the total pressure to the event horizon where stretching of flux
tubes is the most pronounced. Taking all this into account  we choose for
simulation the following expression:

\begin{eqnarray}
P(r)=\frac{c_P}{(r-r_g)^d}, \label{P}
\end{eqnarray}

 where $c_P$ and $d$ are some constants. We run our programme for
 different distributions $P(x^i)$, and the results
 depend weakly on this function. The black hole is such a powerful object
 that qualitatively the behaviour of a string mainly depends on the
 characteristics of the black hole itself (angular momentum, first
 of all) rather than details of pressure distribution. However,
 there are two important restrictions on $P(x^i)$. It should not
 be too big to exceed the gravitational attraction, and it can not be
 too small because magnetic field strength is limited by the total
 pressure (\ref{11}).

{\bf 6.3 String equations}. From the numerical point of view it is convenient
to rewrite the string equations (\ref{10}) in conservative form using the
variational method:

\begin{eqnarray}
& & \frac{\partial }{\partial \tau}
 \left(\frac{Q w}{\rho}x_{i\tau}\right)-
\frac{\partial }{\partial \alpha}
 \left(\frac{\rho}{4 \pi w}x_{i\alpha}\right)=
 \frac{1}{\rho w} \frac{\partial P}{\partial {x^i}} \nonumber\\ [1mm]
& &  \hspace*{2.5cm} +  \frac{1}{2}\frac{\partial g_{jk}}{\partial x^i}
 \left(\frac{Q w}{\rho}x^j_\tau x^k_\tau -
 \frac{\rho}{w}x^j_\alpha x^k_\alpha\right)
 , \label{10a}
\end{eqnarray}

where $x_{i\tau} \equiv g_{ik} x^k_\tau, x_{i\alpha} \equiv g_{ik}x^k_\alpha$.
The equations (\ref{10a}) have an advantage in comparison with (\ref{10}) in
that energy density (\ref{21}) and angular momentum density (\ref{22}) are
explicitly included in the numerical scheme.

Adding the following  equations:

\begin{eqnarray}
\frac{\partial x^i_\alpha}{\partial \tau} & = & \frac{\partial x^i_\tau
}{\partial \alpha},  \label{at}
\\[1mm]
\frac{\partial x^i}{\partial \tau} & = & x^i_\tau
 , \label{xt}
\end{eqnarray}

we can present the string equations in terms of 12-dimensional vectors of state
$W$, flux $F$ and source $S$:

\begin{eqnarray}
\frac{\partial W}{\partial \tau}- \frac{\partial F }{\partial
\alpha}=S, \label{wfs}
\end{eqnarray}

where

\begin{eqnarray}
W & = & \left(\frac{Qw}{\rho}x_{i\tau}, x^i_\alpha, x^i \right), \label{W}
 \\[1mm]
F & = & \left(\frac{\rho}{w}x_{i\alpha}, x^i_\tau, 0 \right), \label{F}
\\[1mm]
S & = & \left(\frac{1}{\rho w} \frac{\partial P}{\partial {x^i}}+
 \frac{1}{2}\frac{\partial g_{jk}}{\partial x^i}
 \left(\frac{Q w}{\rho}x^j_\tau x^k_\tau -
 \frac{\rho}{w}x^j_\alpha x^k_\alpha\right),0, x^i_\tau
 \right).
  \label{S}
\end{eqnarray}

We solved the string equation (\ref{wfs}) numerically using the ordinary
Lax--Wendroff method as well as TVD scheme.

{\bf 6.4 Calculation of density}. The plasma is assumed to be polytropic with
the following  equations of state for gas pressure, internal energy and
enthalpy in dimensionless form:

\begin{eqnarray}
p = c_e \rho^\kappa,\,\ \varepsilon=1+\frac{c_e}{\kappa-1}\rho^{\kappa-1},\,\
w=1+\frac{c_e\kappa}{\kappa-1}\rho^{\kappa-1}. \label{prho}
\end{eqnarray}

Note that the relativistic term $\rho c^2$ is taken into account, and, for
example, the enthalpy per unit mass is equal to $w=c^2+\frac{p}{\rho}$.

The plasma density is obtained from the following transcendental
equation:

\begin{eqnarray}
P(x^i)=c_e \rho^\kappa - \frac{1}{2}\rho^2 g_{ik}x^i_\alpha x^k_\alpha,
\label{rho}
\end{eqnarray}

which is just a definition of the total pressure. For the particular case
$\kappa=2$ the density can be found analytically:

\begin{eqnarray}
\rho=\sqrt{\frac{P(x^i)}{c_e  - \frac{1}{2} g_{ik}x^i_\alpha x^k_\alpha}}.
\label{rho1}
\end{eqnarray}

For the sake of simplicity we will suppose the polytropic index to be 2.

 {\bf 6.5 Initial state}. For the initial moment $\tau=0$ the flux tube is
  assumed to be  elongated along the $x$-axis outside the ergosphere:

\begin{eqnarray}
x^1=c_x \alpha,\,\ x^2=y_0,\,\ x^3=z_0, \label{x0}
\end{eqnarray}

where $c_x, y_0, z_0$ are constants. Then the plasma density is determined from
 equation (\ref{rho1}).

We assume that the initial angular momentum of each element of the string
vanishes, hence

\begin{eqnarray}
\phi_\tau=-\frac{g_{t\phi}}{g_{\phi \phi}}t_\tau. \label{phi0}
\end{eqnarray}

Then the initial velocity $t_\tau$ can be obtained from the gauge condition
(\ref{h5})

\begin{eqnarray}
t_\tau=\frac{1}{w}\sqrt{\frac{g_{\phi\phi}}{g_{tt}g_{\phi
\phi}-g_{t\phi}^2}}. \label{phi1}
\end{eqnarray}

Using equations (\ref{x0}- \ref{phi1}) we can find $W,F$ and $S$ for the
initial moment $\tau=0$.

{\bf 6.6 Boundary conditions}. The length of the flux tube must be sufficiently
long to avoid influence of energy and momentum flux through the edges of the
string. We used free boundary conditions at the edges:

\begin{eqnarray}
\frac{\partial}{\partial \alpha} x^i_\tau=0, \label{bn}
\end{eqnarray}

and controled conservation of the total energy and angular momentum of the
string.

{\bf 6.7 Tortoise variable}. To avoid too big a step along the radial
coordinate $r$ we used the tortoise distance \cite{14}

\begin{eqnarray}
s=-\mbox{ln}(r-r_h), \label{bn}
\end{eqnarray}

where $r_h$ is the radius of the event horizon.

{\bf 6.8 Error control}. The problem under consideration is a stiff one, and at
some stage errors of calculation started to grow exponentially. Therefore it is
important to control accuracy of the numerical scheme. To this end we used the
gauge condition (\ref{h5}) which has not been actually involved in the
numerical method. If the difference

\begin{equation}
w\sqrt{g_{ik}x^i_{\tau}x^k_{\tau}}-1<\varepsilon_0 \label{accu}
\end{equation}

exceeds some level $\varepsilon_0$ we stop the calculations. The value
$\varepsilon_0=0.03$ has been taken in  most of the runs.

{\bf 6.9 Initial parameters}. We present hereafter results of our simulation
for the following parameters:

\begin{equation}
c_e=0.5,\,\ c_P=0.01, \,\ c_x=1, \,\ d=2, \,\ y_0=1.9, \,\ z_0=1.4. \label{ini}
\end{equation}

The number of grid points along the string  $-20<\alpha<20$ is chosen to be
4000. It will be shown in Figures only the central part $-10<\alpha<10$. The
black hole is supposed to be near the extreme rotation $a=.99 M$.
\\

\noindent{\bf 7. Results of numerical simulation}\\

{\bf 7.1 Geometry of the falling flux tube}. From the numerical simulation we
obtain the functions $x^i(\tau,\alpha)$,$x_{i\tau}(\tau,\alpha)$,
$x^i_\alpha(\tau,\alpha)$. The evolution of Boyer-Lindquist coordinates
$t,r,\theta,\phi$ which specify the position of the string as a function of
$\alpha$ for different string times $\tau=\delta, 2\delta,...,7\delta$ is shown
in Figure 1, where $\delta=5$. One can see that the central fragment of the
string is attracted by the hole more strongly compared with the trailing parts.
Hence, this fragment is stretched in the direction of the hole, and
simultaneously the string becomes involved into rotation around the Kerr black
hole. When the leading part encounters the outer boundary of the ergosphere,
the static limit surface, both effects (attraction and rotation) become more
pronounced. Eventually, the flux tube sweeps around the horizon producing a
characteristic spiral structure.

To investigate the string geometry we embedded the string into 3D flat space
using pseudo-Cartesian coordinates \cite{17}:

\begin{eqnarray}\label{xyz}
x & = & \sqrt{r^2+{r_g}^2} \mbox{sin}\theta \mbox{cos}\phi, \nonumber \\[1mm] y
& = & \sqrt{r^2+{r_g}^2} \mbox{sin}\theta \mbox{sin}\phi, \\[1mm]  z & = &
r\mbox{cos}\theta. \nonumber
\end{eqnarray}

The mapping (\ref{xyz}) corresponds to the case $M=0$ (mass of the hole
vanishes), and the Boyer-Lindquist coordinates describe a flat space.

One can observe in Figure 2 how the black hole attracts the string and makes it
spin fast around the event horizon.

It is interesting that the motion of the string inside the ergosphere is rather
complicated: the falling tube simultaneously slowly approaches the event
horizon, quickly rotates, moves from the equator to the cusp region, is
reflected, returns back to the equator and so on (Figure 1,2). Therefore the
spiral structure of the string inside the ergosphere is very complicated.

In the course of time plasma flows into the ergosphere, but no fluid element
can escape the latter.

{\bf 7.2 Energy and angular momentum}. The distribution of the energy density
(first panel) and angular momentum density (second panel) along the falling
tube for the same string times is shown in the second column of  Figure 1.

Our simulations confirm the general conclusions of the previous section based
on conservation laws and differential rotation. The leading fragment of the
falling tube is located closer to the black hole, is more stretched, hence
string braking acts more strongly. As a result this part of the tube has to
rotate more slowly than space (i.e., fiducial observers) which implies that
negative angular momentum has to be generated. However, the angular momentum of
the whole tube which vanishes at the initial moment has to be conserved, and to
compensate for this loss positive angular momentum appears and propagates along
the tube (see Figure 1). In the course of time when the tube enters the
ergosphere this effect becomes more and more pronounced, but the negative
angular momentum is created most efficiently when the tube is deep down near
the horizon.

It is apparent  that such a behaviour of angular momentum should have some
consequences also in the behaviour of the tube energy, but surprisingly it
turns out that the angular momentum distribution is nearly proportional to the
energy distribution along the string which can be seen by comparing the two
practically identical pictures in Figures 1. This circumstance allows to
include energy  as a last term in the chain of physical processes related to
the leading fragment of the tube:

\begin{eqnarray}\label{scheme}
& & \mbox{Stretching} \rightarrow \mbox{Braking} \rightarrow \nonumber \\& &
\mbox{Negative angular momentum} \rightarrow \mbox{Negative energy}.
\end{eqnarray}

The physical reason for this effect is that negative energy is created inside a
narrow layer near the horizon where the time-coordinate is approximately
proportional to the rotation (compare panels for $t(\tau,\alpha)$ and
$\phi(\tau,\alpha)$ in Figure 1).

With regards to energy problem it was not clear in advance whether or not  the
flux tube gets negative energy inside the ergosphere, but our simulations show
that in fact it does. As one can see in Figure 1, the energy density
progressively decreases from the initial value $\sim 1$ to less than $-20$ at
$\tau=30$. Simultaneously plasma density increases from $\sim .05$ to about
$180$, and magnetic energy density $B=h^i h_i/2$ increases from $\sim .05$ to
$\sim 100$ (third and fourth panels in the second column in Figure 1).

The negative energy starts to be created approximately at $\tau=10$ (see Figure
3) and then decreases linearly with string time. As a result, the part of the
tube with positive energy gets extra energy from the black hole, and ends up
with  double the energy $E \sim 80$ at $\tau=30$ in comparison  with the total
initial energy of the string $E_0 \sim 40$.

{\bf 7.3 Energy confinement}. The  negative energy as well as the negative
angular momentum are created in the narrow layer very near to the event horizon
in the range $r_g=.51<r<.52$, and moreover  most of the compensated positive
energy and angular momentum is also located in this region (see Figure 4a).

There is also a wave of positive $E$ and $L$ (see Figure 4b) which propagates
towards the outer boundary of the ergosphere located at $r \sim 1 $.
Unfortunately, the speed of the wave is approximately equal to the velocity of
the plasma entering the static limit surface, and as a result the wave can not
leave the ergosphere. Both negative and most of the positive energy turns out
to be localized near the black hole, and there is some kind of energy
confinement inside the ergosphere.

Nevertheless, this confinement is not absolute. The wave with $E>0$ can still
carry some small amount of energy out of the ergosphere. For example, the
energy density near the static limit surface $r=1$ at $\tau=30$ is double  the
initial value (see Figure 4b), and very slowly increases with string time.
Therefore the tube can extract energy from the ergosphere but at a relatively
low rate compared to the negative energy creation rate. The paradox of the
situation is that the falling string can effectively generate the extra
positive energy by taking it from the rotating black hole, but on the other
hand can not effectively transport the extra energy through the outer boundary
of the ergosphere.

We  run simulations with different initial parameters  of the string such as
position, total pressure behaviour, etc., but the main result stays the same:
the extra energy of the falling tube is created near the event horizon at a
much higher rate than that for the extraction of energy out of the
ergosphere.\\

\noindent{\bf 8. Test particle - flux tube comparison}\\

It is interesting to compare the behaviour of a test particle with that of a
tube/string. We noted previously that the formal expressions for the energy and
angular momentum (\ref{21}- \ref{22a}) are similar up to a factor $Q/w\rho$. We
can therefore identify a link between a test particle and a string. On the
other hand there is also a major difference between them. In particular,  test
particles can be regarded  as independent entities, while fluid elements
closely are coupled to each other. Thus, $E$ and $L$ are conserved separately
for each particle, but for the string the conservation laws hold for the tube
as a whole, and we can not apply them to the individual fluid elements. This
statement has several important implications:

i. The Penrose mechanism \cite{3}, which is based on the test particle
approach, requires at least two particles for exchange and transmission of
energy and angular momentum. In the string approach, in contrast, a single flux
tube is sufficient to extract energy from the black hole.

ii. Generally, the trajectories of test particles and flux tubes differ
significantly. Although the trajectories may start off the same initially if
the magnetic field involved is small, eventually  the effects of the magnetic
field and associated stresses lead to completely different behaviour. In our
simulations, for example, we did not find any signatures of stable orbits.

iii. It is known that a test particle takes an infinite time to reach the
horizon relative to an observer at infinity, even though from the point of view
of the particle itself this take only a finite time and can even be rapid.

For the string, on the other hand, longer time intervals imply increasingly
stronger magnetic stresses, as the string gets tightly wound around the event
horizon. Thus, an infinite time interval would probably lead to an infinite
amount of tension, and we would not expect stress to build up to this extent.
To get round this problem, we need to invoke an additional process. Since the
leading part of the string develops a loop, which gives rise to an antiparallel
magnetic field configuration (see Figure 2), it would seem natural to suggest
that reconnection could fulfill the role of this additional process.
Reconnection is able to cut the fragment of the tube with negative energy,
which corresponds to this loop, and thus release stresses which are building
up, similar to what happens in current sheet configurations in the
magnetospheric and solar context \cite{5,18,19}. In the case of the
solar-terrestrial interaction, we know that flux tubes filled with solar plasma
flow around the obstacle formed by the Earth's magnetic field. This is
accomplished through reconnection of magnetic field lines at the boundary
between the solar wind and the magnetosphere. Reconnection releases the
stresses which would otherwise build up as the solar flux tubes hang and drape
around the magnetosphere \cite{18}. Similarly, a flux tube which encounters a
black hole gets wound up around the event horizon, as demonstrated by our
simulations. The only way, it seems, to release the resulting stresses and
capture some of the energy inside the ergosphere to transmit it to the outside
world, is through reconnection. This results in part of the flux tube with
negative energy falling into the black hole, and the other part can then
transfer some of the positive energy extracted from the black hole outside the
ergosphere. The mechanism we have describe here whereby a magnetic flux tube is
captured by a black hole and then extracts and releases some of the energy
contained in the rotation of the black hole to the outside world may therefore
be viable  one to explain such phenomena as astrophysical jets \cite{1}.\\

\noindent{\bf 9. Discussion and conclusion}\\

The main aim of this paper is to present an example of how a falling flux tube
can extract energy from a Kerr black hole. According to our numerical
simulations, due to stretching and braking, the energy of the leading part of
the tube becomes negative. Since the total energy of the string is conserved,
the rest of the tube gets extra energy from the hole, i.e., extracts spinning
energy from the latter. So, our simulations confirm in general the prediction
of \cite{2} concerning magnetic flux tube - Kerr black hole coupling, but some
unexpected features cropped up.

For a sufficiently strong rotation of the hole and for a sufficiently strong
initial magnetic field of the falling tube, the latter  can effectively
generate negative energy, however the fragments of the string with both
negative and compensated positive energy are localized inside the ergosphere.
The energy extracted from the black hole  is confined inside the ergosphere and
almost does not show up outside, and probably can not produce such a phenomena
as relativistic jets from the AGN.

There are several possibilities for removing the spinning energy from the hole
more effectively.

First of all, so far we have only studied the initial phase of the flux tube
motion inside the ergosphere of a Kerr Black hole. We were able to run
programmes only for relatively short string-time intervals $\tau < 30$ before
errors of the numerical scheme exceeded the limit $3\%$. This corresponds, for
example, for a black hole of a solar mass to a time duration of only $0.3$ ms
which is far well outside astronomical scales. The relativistic flux tube is a
strongly non-linear system, and theoretically we can not avoid the possibility
that energy will be extracted much more effectively for later times. On the
other hand, our results show no signatures of such a behaviour, and it seems
reasonable to conclude that the negative energy creation rate will always be
much bigger than that of energy extraction from the ergosphere. Of course, this
is only true if we do not  invoke some additional processe.

An alternative is to consider the behaviour of the total pressure, which we
have to define a priori in the string approach. Certainly, we were not able to
try all possible variants of the total pressure distribution, and some of them
may lead to a more powerful process of energy removal from a black hole. There
also might be a scenario in which the whole problem is essentially
time-dependent. We can imagine that accreting magnetized plasma which consists
of many flux tubes, can produce a pressure which increases with time due to
period of a long rotation around the horizon which eventually pushes the plasma
out of  the ergosphere in a bursty explosive-like regime. We can not answer all
these questions right now, and they left for furure studies.

Finally, the most promising process which can produce relativistic jets seems
to be magnetic reconnection. Our conclusion is based on the investigation of
the Earth's magnetosphere, where we have the benefit of direct spacecraft
measurements. It turns out that the energy of the magnetospheric substorm comes
from the magnetic energy stored in the magnetotail which is released in an
explosive way by virtue of reconnection producing narrow flows of accelerating
plasma \cite{18}. The reconnection events at the magnetopause \cite{18,20}
(boundary of the magnetosphere) and solar flares \cite{19} present other
examples of this exlosive energy release, when magnetic energy is first
accumulated and then reconnection converts it into kinetic and internal energy
to produce jets of accelerated plasma.

Therefore, we believe that our simulations reproduce the initial stage
corresponding to the storage of the magnetic energy inside the ergosphere,
which then creates conditions for a bursty regime  of energy release involving
reconnection. Inside the ergosphere parts of the flux tube with oppositely
directed magnetic field are very close to each other (see Figure 2) and
therefore reconnection \cite{21,22} can come into play. The fragment of the
falling flux tube with negative energy is absorbed by the hole, but the other
one receives extra energy from the hole, thus producing a relativistic jet.
Physically, plasma is accelerated to relativistic velocities by slow shocks
which propagate along the tube and release Maxwellian tension. It is very
important to note that acceleration does not happen in one instant, but it
occurs continuously as the shocks propagate along the magnetic field. So, the
initial quasi-stationary regime generates Maxwellian tension along the string,
and subsequently reconnection releases this tension to give rise to a
relativistic jet.

Although reconnection can occur at any point near the black hole, only when it
occurs inside the ergosphere can energy be extracted from the hole, and thus
accelerate plasma to the relativistic velocities needed for cosmic jets. The
stability of cosmic jets are explained in our model in terms of the
relativistic velocities of the plasma and the spiral structure of the magnetic
field.

It is worthwhile discussing the string model in relation to the
Blandford-Znajek  mechanism \cite{1,23}, which currently seems to be the most
popular one. The string process is inherently three dimensional and
time-dependent because it is based on differential rotation even for the
quasi-stationary regime, although the magnetic field structure in the final
stage may be nearly axisymmetric. It is well known that for a purely
steady-state axisymmetric configuration, the flux surface must have rigid
rotation \cite{24} which is not true for the string mechanism. There is only a
tendency to rigid rotation, which is clear from the redistribution of the
angular momentum, i.e., the leading part of the falling tube spins more slowly,
the trailing part faster. But to establish exactly rigid rotation takes an
infinite time, which can not  happen in reality. Therefore, the string
mechanism differs from the Blandford-Znajek one based on steady-state and
axisymmetrical patterns. Besides, the Blandford-Znajek process needs a magnetic
field embedded in the hole's event horizon while the flux tube in the string
approach does not reach the horizon at all so far, which emphasizes  the
difference even more.

Nevertheless, there are also common features for both mechanisms. First of all
the main idea of the Blandford and Znajek mechanism \cite{1,23} is that
magnetic forces are the most appropriate to couple the black hole's spin to
external matter. In other words, these mechanisms are the non-local variants of
the original Penrose process \cite{3} which make use of electromagnetic effects
to remove spin energy from a Kerr black hole. Therefore we believe that future
development could reveal that the Blandford-Znajek mechanism and the string
model are just two different ways for describing the same physical process,
even though they now look  rather different.

There is a direct generalization of the scheme described above to cosmic
strings. It is a fact that formally the equations of motion for ordinary cosmic
strings generated by the Nambu-Goto action \cite{16} are the same as those for
the flux tubes (\ref{10}) if we replace the factors $Qq/\rho$ and $\rho/4 \pi
q$ with an energy per unit length $\mu$, estimated as $10^{20} kg/m$. The only
important difference is the gauge conditions, but they do not influence the
energy and momentum conservation laws. There is an even  closer analogy with
the current-carrying strings, for which the energy per unit length and the
string tension are in general different \cite{25}. Therefore, we can expect
extraction of spin energy of the hole by a cosmic string in the form of a
spiral wave, and then reconnection can release the string tension to produce a
fast moving part of the cosmic string.

Accretion with differential rotation leads to a general pattern with a spiral
structure of the magnetic field and a plasma flow which eventually may result
in jet formation by virtue of reconnection. The prediction of such a pattern is
based just on the application of energy and angular momentum conservation for
the flux tube, and there is nearly no dependence on the initial configuration
of the plasma flow and magnetic field; the only restriction is that they should
not be axisymmetric.

We believe that the string mechanism described above can be used to develop
model of active galactic nuclei.\\

{\sl Acknowledgements}: We are grateful to I. V. Kubyshkin and  R. P. Rijnbeek
for helpful discussion. Part of this work was done while VSS, SAD, and IBI were
on research visit to Graz. This work is partially supported by the Russian
Foundation of Basic Research , grant No. \mbox{01-05-64954}, by INTAS-ESA,
grant No. 99-01277, and by the programme INTERGEOPHYSICS from the Russian
Ministry of Higher Education. SAD was suported by INTAS, grant No. YSF-80 and
by the Nansen Center. Part of this work is supported by the Austrian ``Fonds
zur F\"orderung der wissenschaftlichen Forschung'', project P13804-TPH. We
acknowledge support by the Austrian Academy of Sciences, ``Verwaltungsstelle
f\"ur Auslandsbeziehungen''.\\

\newpage

\setcounter{figure}{0} 

\begin{figure}
\caption{Numerical results for a flux tube falling onto a Kerr black hole  as
functions of the mass parameter $\alpha$. Distributions of different parameters
along the string are shown for the string times $\tau=\delta,
2\delta,...,7\delta$ where $\delta=5.$ Left column: distributions of the
Boyer-Lindquist coordinates $t,r,\theta,\phi$. Right column: distribution of
the density of the energy, of the angular momentum, of the plasma, and of the
magnetic energy.}
 \label{fig:1}
\end{figure}

\begin{figure}
\caption{Visualization of the flux tube geometry by mapping Boyer-Lindquist
coordinates to pseudo-Cartesian coordinates for the initial stage $\tau=\delta,
2\delta,...,9\delta$, where $\delta=1.2$ (top) and for the late string time
$\tau=30$ (bottom). The fragment of the string with negative energy is marked
with red points.}
 \label{fig:2}
\end{figure}

\begin{figure}
\caption{Behaviour of the total (solid), negative (dash-dotted), and positive
(dotted) energy of the string in the course of string time.}
 \label{fig:3}
\end{figure}

\begin{figure}
\caption{Distribution of the energy density of the string as a function of $r$
(a) in the narrow layer near the horizon  and (b) in the vicinity of the
ergosphere. The fragment of the tube with $\alpha<0$ is marked with points.}
 \label{fig:4}
\end{figure}

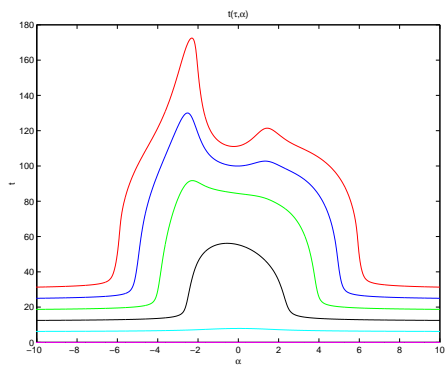
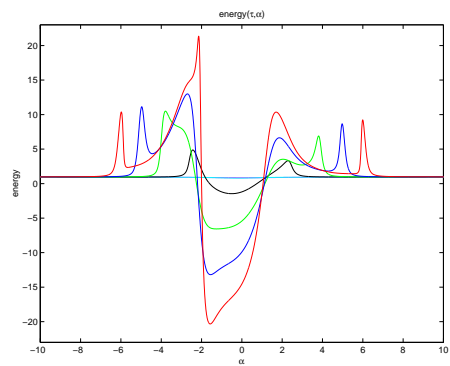
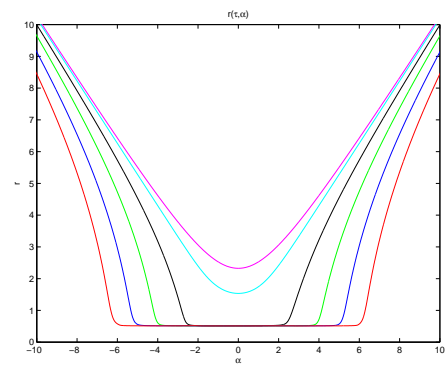
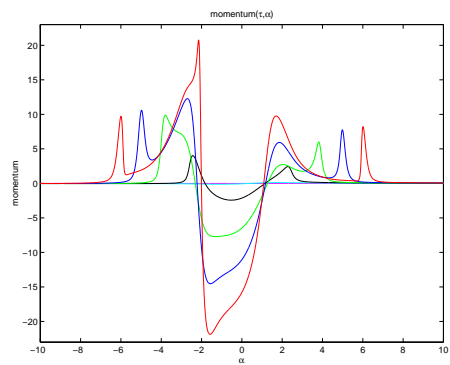
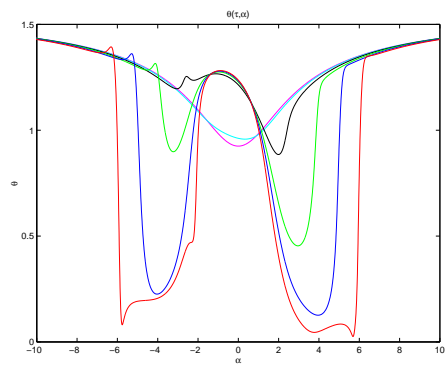
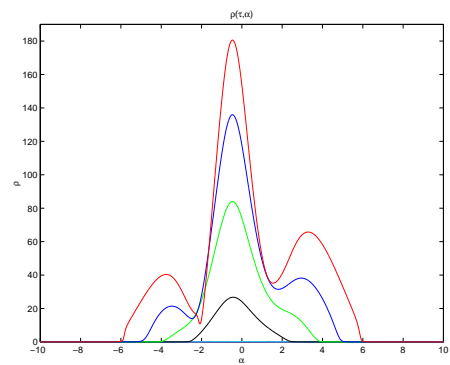
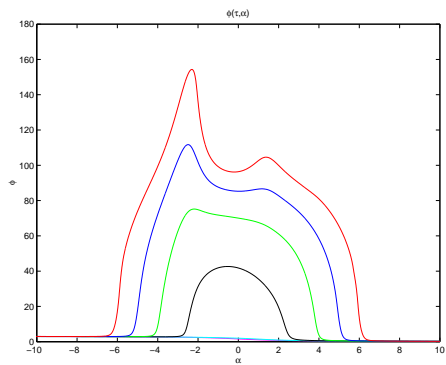
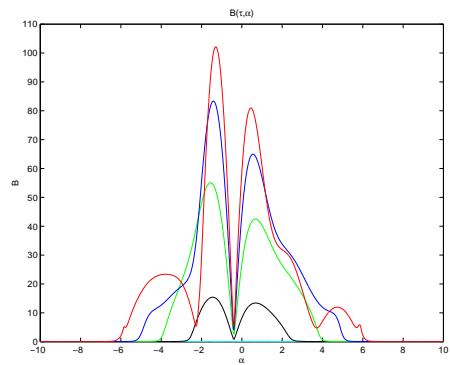
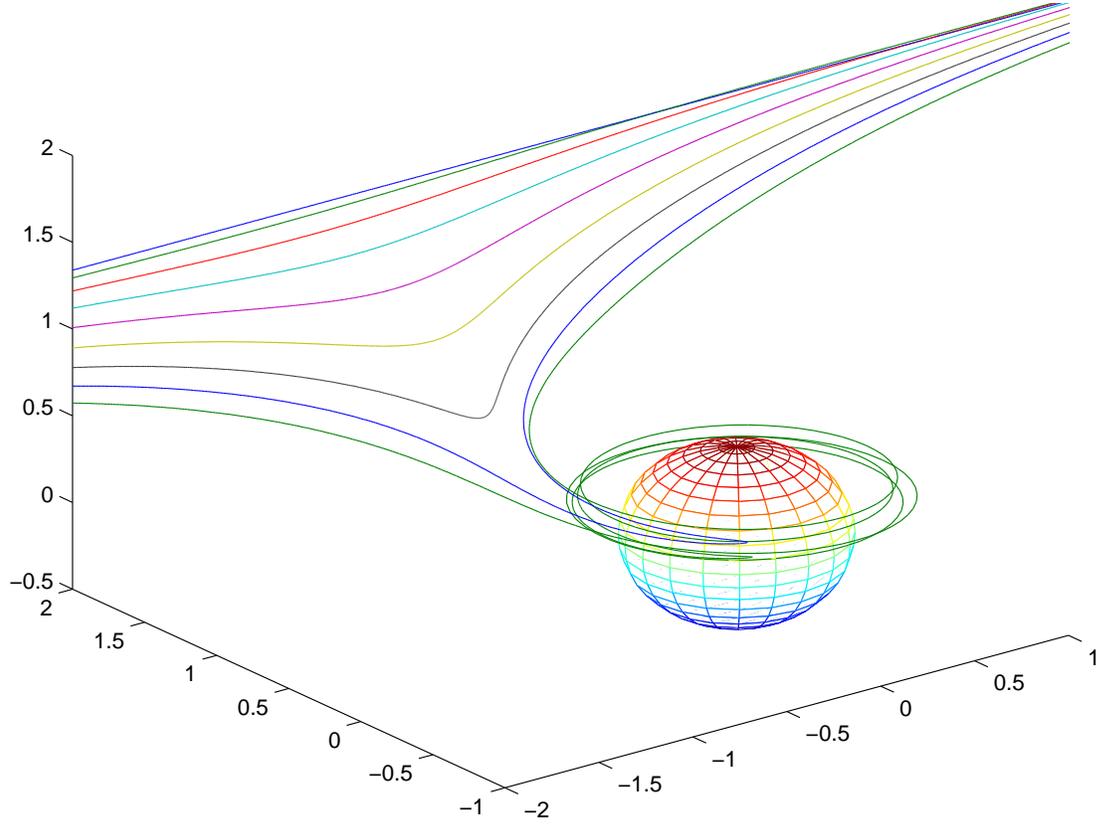
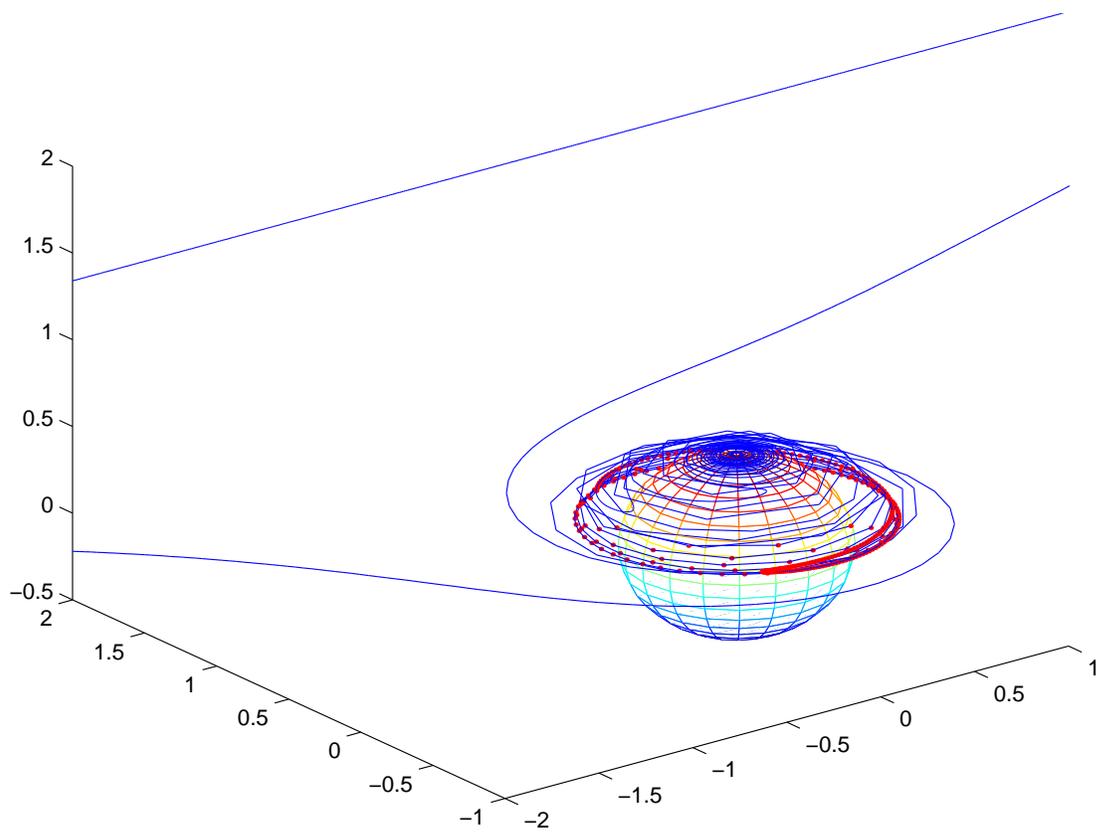
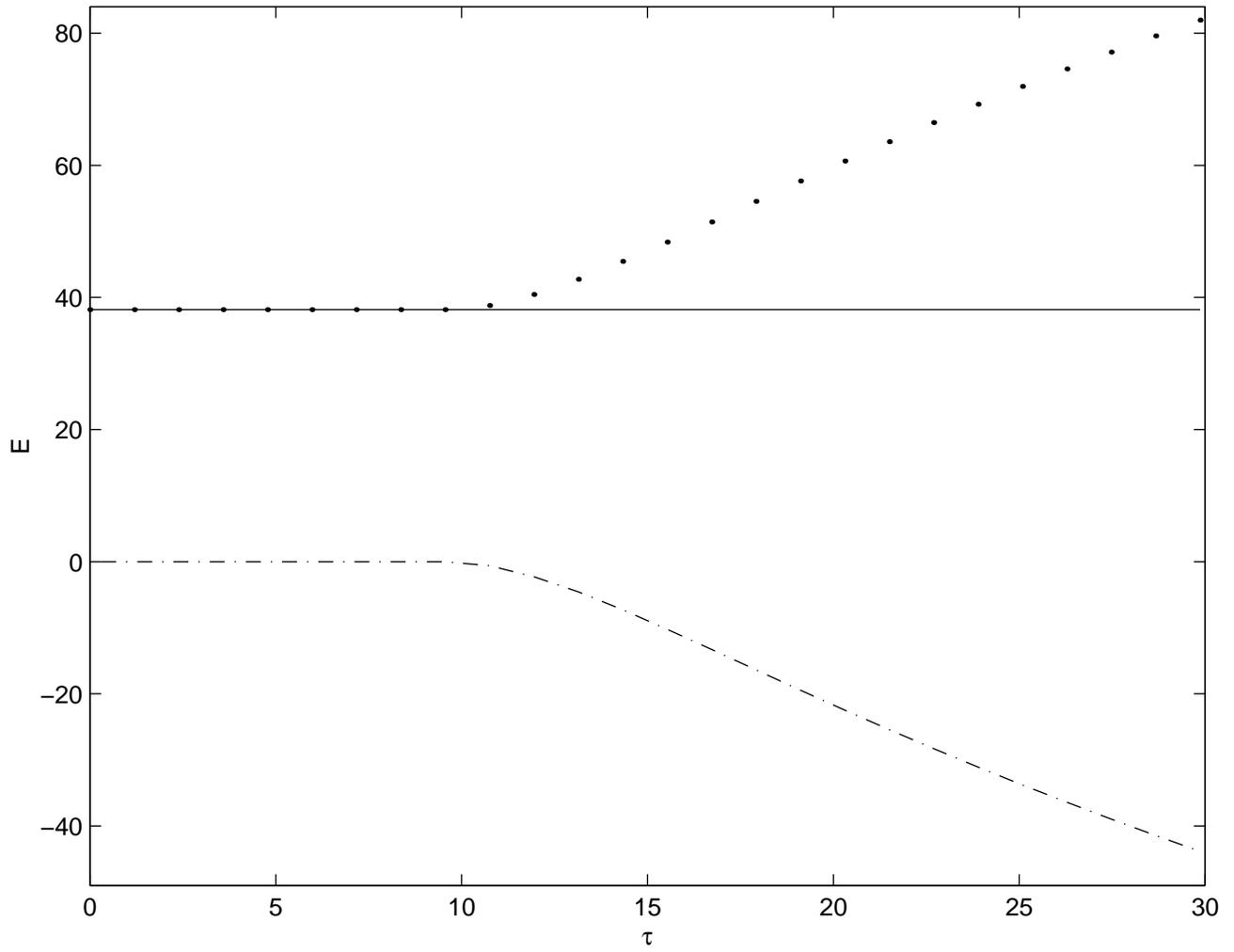
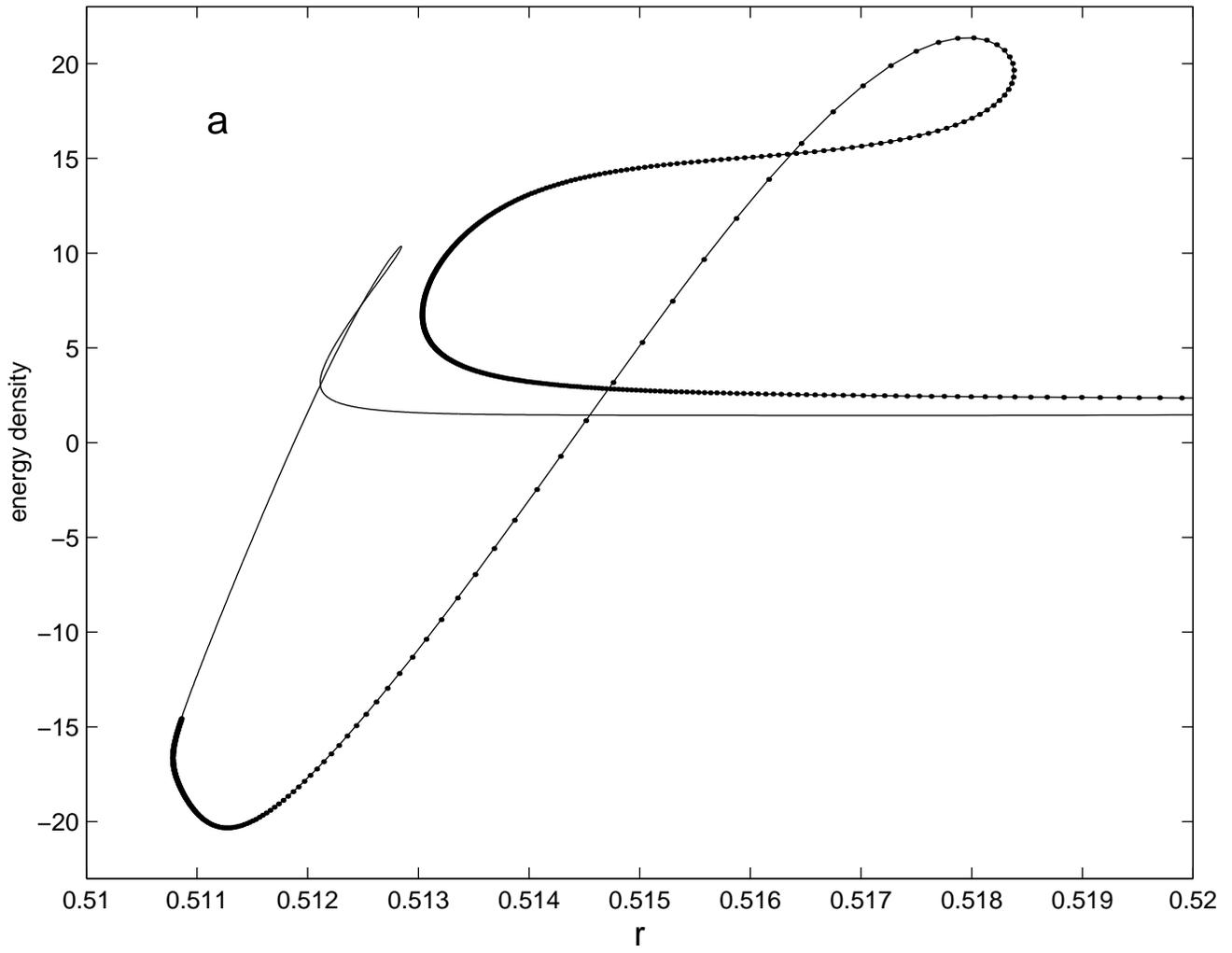
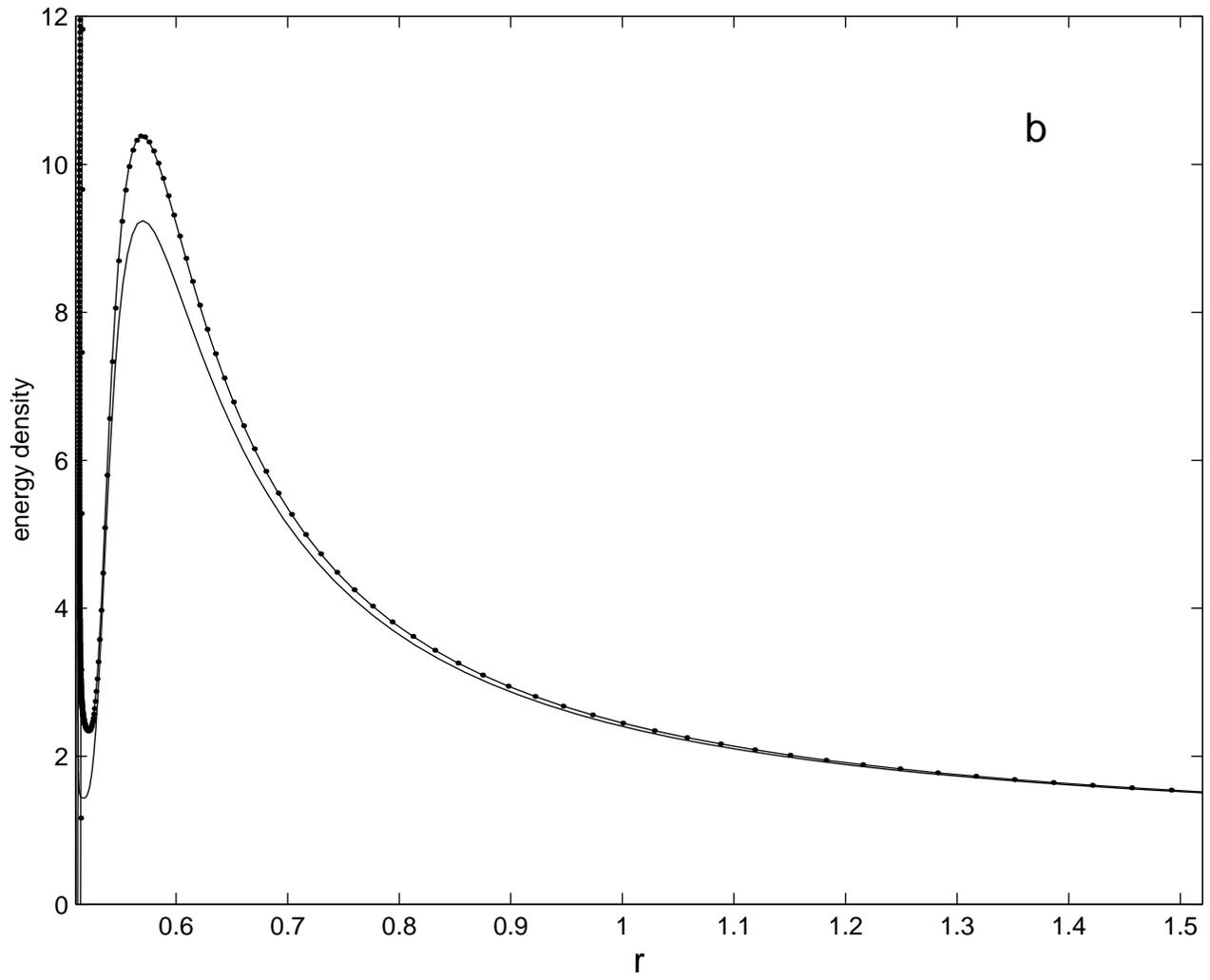
\end{document}